\documentclass[letterpaper, 10 pt, conference]{ieeeconf}  
\IEEEoverridecommandlockouts 
\pdfoutput=1
\usepackage[noadjust]{cite}
\usepackage{graphicx}
\usepackage{subcaption}
\usepackage{multirow}
\usepackage{url}
\usepackage{amsfonts}
\usepackage{amsmath}
\usepackage{bm}
\usepackage{color}
\usepackage{float}
\usepackage[LY1]{fontenc}
\usepackage{comment}
\usepackage{array}
\usepackage{afterpage}
\newcolumntype{P}[1]{>{\centering\arraybackslash}p{#1}}

\usepackage[ruled,vlined]{algorithm2e}
\usepackage{algpseudocode}   
\newcount\DraftStatus  
\newcommand{\draftnote}[2]{
	\ifnum\DraftStatus=1
	\marginpar{
		\scriptsize\raggedright
		\hbadness=10000
		\def\baselinestretch{0.9}
		\textcolor{#1}{\textsf{\hspace{0pt}#2}}
	}
	\fi
}

\DeclareMathSizes{10}{9}{7}{5}

\title{\LARGE \bf
A New Non-Negative Matrix Co-Factorisation Approach\\ for Noisy Neonatal Chest Sound Separation}
\author{Ethan Grooby, Jinyuan He, Davood Fattahi, Lindsay Zhou, Arrabella King, Ashwin Ramanathan, \\Atul Malhotra, Guy A. Dumont, ~\IEEEmembership{Life Fellow,~ IEEE}, Faezeh Marzbanrad,~\IEEEmembership{Senior Member,~IEEE} 
\thanks{E. Grooby, J. He and F. Marzbanrad are with the Department of Electrical and Computer Systems Engineering, Monash University, Melbourne, VIC, Australia.}
\thanks{L. Zhou, A. King, A. Ramanathan and A. Malhotra are with Monash Newborn, Monash Children’s
Hospital and Department of Paediatrics, Monash University, Melbourne, Australia.}
\thanks{G. Dumont is with the Department of Electrical and Computer Engineering, University British Columbia, Vancouver, BC, Canada and with the BC Children's Hospital Research Institute, Vancouver, BC, Canada.}
\thanks{E. Grooby  acknowledges the support of MIME-Monash Partners-CSIRO sponsored PhD research support program. A. Malhotra research is supported by the Kathleen Tinsley Trust and a Cerebral Palsy Alliance Research Grant. F. Marzbanrad acknowledges the support of Advancing Women's Research Success Grant program.}

\thanks{email: ethan.grooby@monash.edu}
}
\begin{document}
\maketitle
\thispagestyle{empty}
\pagestyle{empty}

\begin{abstract}
Obtaining high quality heart and lung sounds enables clinicians to accurately assess a newborns cardio-respiratory health and provide timely care. However, noisy chest sound recordings are common, hindering timely and accurate assessment. A new Non-negative Matrix Co-Factorisation based approach is proposed to separate noisy chest sound recordings into heart, lung and noise components to address this problem. This method is achieved through training with 20 high quality heart and lung sounds, in parallel with separating the sounds of the noisy recording. The method was tested on 68 10-second noisy recordings containing both heart and lung sounds and compared to the current state of the art Non-negative Matrix Factorisation methods. Results show significant improvements in heart and lung sound quality scores respectively, and improved accuracy of 3.6bpm and 1.2bpm in heart and breathing rate estimation respectively, when compared to existing methods. 

\end{abstract}

\section{Introduction}
Accurate and timely assessment for signs of serious health problems such as cardio-respiratory diseases is an essential requirement to provide care to newborns \cite{Newborns19:online}. Recording chest sounds with a stethoscope is a common and simple method to obtain such information. In recent times, the availability of digital stethoscopes for neonates has attracted several studies \cite{ramanathan2019digital,kevat2017digital,zhou2020acoustic,ramanathan2020assessment}. However, higher level of noise in neonatal intensive care in comparison to adult and paediatric wards, has resulted in poor quality chest sound recordings and inaccurate assessment. For instance, estimation of heart rate and breathing rate from low quality signals has been shown to be error-prone \cite{grooby2020neonatal}. Noise interference in chest sounds can include external and background noise \cite{lahav2015questionable}, heart or lung sound acting as noise for one another, other internal sounds such as bowel sounds, gastric reflux and air swallow, and respiratory support equipment noise. Overall, it is essential to reduce external, internal and respiratory support noises, and separate heart and lung sounds prior to any assessment and diagnosis.  

Denoising and sound separation methods to obtain high quality heart and lung sounds can be broken up into multi-channel and single channel methods. In multi-channel methods, a reference signal such as electrocardiogram or secondary microphone placed either to capture external noise and/or secondary chest recording is used. Utilising this reference signal, adaptive filtering methods, direct removal and interpolation or Non-negative Matrix Factorisation (NMF) can be used to remove this signal from the primary recording \cite{nersisson2017heart, gnitecki2007separating, shah2014blind}. However, as typically the reference and primary signals are both mixtures of the desired heart and lung sounds along with the other noises, these are not ideal approaches. Blind source separation and independent component analysis methods address this problem by separating the reference and primary signals into their primary components which are assumed to be heart and lung sounds \cite{nersisson2017heart}. Overall these methods require additional sensors which are not always accessible and feasible to implement. 

For single source sound separation and denoising, current methods have proven only partially effective. A common approach is to first obtain a reference heart sound signal through methods such as singular spectrum analysis, wavelet denoising and adaptive line enhancement, and then perform similar processes as in multi-channel methods \cite{ghaderi2011localizing, gnitecki2007separating, tsalaile2007separation}. 
However, the accuracy of obtaining reference heart sounds is dependent on low noise content and fixed heart-based parameters based on adults. 
For newborns, noisy chest sound recordings are common and heart sound properties differ from adults, making these approaches not generalisable nor adaptable from recording to recording. Other methods include standard frequency filtering, modulation filtering which involves standard frequency filtering on the frequency domain, as opposed time domain representation of the recording and multi-resolution wavelet decomposition and reconstruction \cite{nersisson2017heart,falk2008modulation, hossain2003overview}. Standard frequency filtering is limited due to frequency overlap with heart, lung and noise sources, whereas modulation filtering and multi-resolution wavelet decomposition and reconstruction are parameter dependent and separate the recording into only heart and lung, without removing other noise sources. 



Three key contributions are presented in this paper. First, a new NMCF approach focused on obtaining high quality heart and lung sounds specifically for the newborn population is proposed. Second, we incorporate a noise component in the NMCF model, to separate the sounds into not only heart and lung sounds but also the noise. Finally, the method is assessed using real-world noisy neonatal chest sounds with heart and lung signal quality and heart and breathing rate accuracy. 


The rest of this paper is organised as follows. Section \ref{sec:background} reviews existing work on NMF and general formulation of proposed NMCF method. Section \ref{sec:methods} presents details implementation of the methods their evaluation. Results and discussion are provided in sections \ref{sec:results} and \ref{sec:discussion}. Section \ref{sec:conclusion} concludes the whole work.

\section{Background}
\label{sec:background}
\subsection{Non-Negative Matrix Factorisation}
Non-negative matrix factorisation decomposes a given non-negative matrix $V\in\Re^{F \times T}_+$ into two non-negative matrices $W\in\Re^{F \times K}_+$ and $H\in\Re^{K \times T}_+$ (Eq. \ref{eq:V}), where $K<min(R,T))$ and E represents the reconstruction error between V and WH. 

\begin{equation}
\label{eq:V}
\begin{split}
V & =WH + E\\ 
& =\Lambda + E
\end{split}
\end{equation}

In denoising and sound separation, V represents the magnitude of time-frequency representation of the recording mixture. 
The weight matrix, W contains the basis vectors $w_1$ to $w_K$ that represent the spectral pattern of different types of signals sources (e.g. heart, lung and noise) or their sub-components and H, the activation matrix ($h_1$ to $h_k$) represents when the signal sources occur during a particular time frame. These sub-components can be combined such that the first set of components (1 to $b_h$), second set of components ($b_h+1$ to $b_h+b_l$) and third set of components ($b_h+b_l+1$ to $b_h+b_l+b_n$) represent heart ($V_h=W_hH_h$), lung ($V_l=W_lH_l$) and noise respectively ($V_n=W_nH_n$) (Eq.\ref{eq:WH}). Where $b_h,b_l,b_n$ are the total number of components used to represent heart, lung and noise with $b_h+b_l+b_n=K$.    
\begin{equation}
\label{eq:WH}
\begin{split}
WH & = [w_1,w_2,...,w_K][h_1;h_2;...;h_K]\\ 
& = [w_1,...,w_{b_h}|w_{b_h+1},...,w_{b_h+b_l}|w_{b_h+b_l+1},...w_K]\\
&[h_1;...;h_{b_h}|h_{b_h+1};...;h_{b_h+b_l}|h_{b_h+b_l+1};...;h_K]\\
& =[W_h|W_l|W_n][H_h;H_h;...;H_n]\\
& =W_hH_h + W_lH_l + W_nH_n\\
& \approx V_h + V_l +V_n\\
Where: &\quad h=Heart, l=Lung, n=Noise
\end{split}
\end{equation}

In the general unsupervised scenario, both W and H are optimised during test time to minimise the cost function D (Eq.\ref{eq:cost}). Most popular cost functions as shown in Eq.\ref{eq:d} are; $\beta =0$ as Ikaura-Saito distance, $\beta =1$ as Kullback-Leibler divergence and $\beta =2$ which yields Euclidean distance. For heart-lung sound separation, Kullback-Leibler divergence has generated the best results \cite{lin2013blind, canadas2017non}. 

\begin{equation}
\label{eq:cost}
W_{opt},H_{opt}=\min\limits_{W,H}D_\beta(V|WH)
\end{equation}

\begin{equation}
\label{eq:d}
D_\beta(x|y)=\begin{cases}
			\frac{(x^\beta-y^\beta-\beta y^{\beta-1}(x-y))}{\beta(\beta-1)}, & \text{if $\beta \in \Re \backslash \{0,1\}$}\\
            x(log(x)-log(y))+(y-x), & \text{if $\beta=1$}\\
            \frac{x}{y}-log(\frac{x}{y})-1, & \text{if $\beta=0$}\\
		 \end{cases}
\end{equation}

A sparsity penalty on the activation matrix H is typically added to enable more detailed decompositions both temporally and spectrally, while ensuring only a small set of meaningful basis vectors are active at a single time frame (Eq.\ref{eq:sparsity}) \cite{le2015sparse}. Overall, this enables a finer level of sound separation. The sparsity penalty is calculated based on the L1-norm of H and $\mu$ controls the importance of the sparsity constraint.

\begin{equation}
\label{eq:sparsity}
\begin{split}
W_{opt},H_{opt} & =\min\limits_{W,H}D(V|\hat{W}H)\\ 
Where: &\quad D(V|\hat{W}H) =D_\beta(V|\hat{W}H) + \mu||H||_1,\\
 &\quad \hat{W}= [\frac{w_1}{||w_1||},\frac{w_2}{||w_2||},...,\frac{w_K}{||w_K||}]
\end{split}
\end{equation}

Based on the cost functions in Eqs. \ref{eq:cost} and \ref{eq:sparsity}, the multiplicative update rule for W and H are shown in Eqs. \ref{eq:hupdate} and \ref{eq:wupdate} respectively. Note that division and $\otimes$ refer to element-wise division and multiplication. 

\begin{equation}
\label{eq:hupdate}
\begin{split}
H &\leftarrow H \otimes \frac{\hat{W}^T(V\otimes\Lambda^{\beta-2})}{\hat{W}^T\Lambda^{\beta-1}+\mu}, \Lambda = WH\\
H & \leftarrow H \otimes \frac{H_{num}(V,W,H)}{H_{dem}(V,W,H)}
\end{split}
\end{equation}

\begin{equation}
\label{eq:wupdate}
\begin{split}
W & \leftarrow \hat{W} \otimes\\ &\frac{(\Lambda^{\beta-2}\otimes V)H^T+\hat{W} \otimes (11^T(\hat{W} \otimes (\Lambda^{\beta-1}H^T)))}
{\Lambda^{\beta-1}H^T+\hat{W} \otimes (11^T(\hat{W} \otimes ((\Lambda^{\beta-2}\otimes V)H^T)))}\\
Where: &\quad \Lambda = WH, \quad 1 =\text{length F column vector of ones}\\
W & \leftarrow W \otimes \frac{W_{num}(V,W,H)}{W_{dem}(V,W,H)}
\end{split}
\end{equation}

\subsection{Single-Source NMF Existing Methods}
For denoising and heart and lung sound separation, the majority of past works remove the noise component using standard frequency filtering or assume no noise is present, and then utilise NMF to separate heart and lung sounds \cite{lin2013blind, shah2014blind, canadas2017non}. 

Two methods are adapted and tested, both methods blindly decompose the mixture into numerous sub-components and then cluster all components into either heart or lung based on spectral or temporal criteria. In Shah et al. method, initial selection of reference basis vectors $W_h^{ref}$ and $W_l^{ref}$, was originally based on the sub-component with strongest power below 100Hz and above 300Hz frequency regions for heart and lung \cite{shah2014blind}. These bands were modified to be 50-250Hz and 250-1000Hz frequency regions, as this matches neonatal heart and lung properties respectively, more closely. Canadas-Quesada et al. utilised three clustering criteria, namely, spectral correlation with clean heart sound database, temporal correlation with detected S1 and S2 heart peaks and frequency roll-off \cite{canadas2017non}. For spectral correlation calculations, we used 10 high quality neonatal heart sounds as defined in section \ref{sec:sqi} for the reference database, as opposed to adult heart sound recordings used in the paper \cite{canadas2017non}. As heart peak detection for temporal correlation used specific parameters related adult heart sounds, this was modified to use a hidden Markov model with heart rate in the range of 70-220bpm, as this provides more accurate results for newborns \cite{grooby2020neonatal}. To avoid using parameter based cutoffs for the clustering criteria which have not been adapted for newborn population, instead, we obtained the sum of the normalised three criteria and top scoring 55 components were assigned to heart and remaining 64 to lung \cite{grooby2020neonatal}. Finally, the more appropriate sparsity implementation (Eqs. \ref{eq:hupdate} and \ref{eq:wupdate}) was used in the implementation of Canadas-Quesada et al. proposed NMF method. 

\subsection{Proposed NMCF Method}

In NMCF, instead of having a training and test phase, which occurs in supervised (Eq.\ref{eq:nmcfs}) and semi-supervised (Eq.\ref{eq:nmcfsemi}) NMF, the matrix basis matrix W is optimised simultaneously with sound separation. This method enables more efficient sound separation as the mixture recording can also contribute to the training of W \cite{de2020wheezing}.

\begin{equation}
\label{eq:nmcfs}
\begin{split}
W_{opt},H_{mopt} & =\min\limits_{W,H}(D(V_m|\hat{W}H_m)+D(V_h|\hat{W_h}H_h)\\
& +D(V_l|\hat{W_l}H_l))\\
Where: & \quad H_m = [H_{mh};H_{ml}], \hat{W}=[\hat{W}_{h},\hat{W}_{l}]\\
\end{split}
\end{equation}

\begin{equation}
\label{eq:nmcfsemi}
\begin{split}
W_{opt},H_{mopt} & =\min\limits_{W,H}(D(V_m|\hat{W}H_m)+D(V_h|\hat{W_h}H_h))\\
Where: &\quad  H_m = [H_{mh};H_{ml}], \hat{W}=[\hat{W}_{h},\hat{W}_{l}]\\
\end{split}
\end{equation}

\begin{figure*}[t]
\centering
\includegraphics[scale=0.15,trim={0cm 0cm 0cm 0cm}]{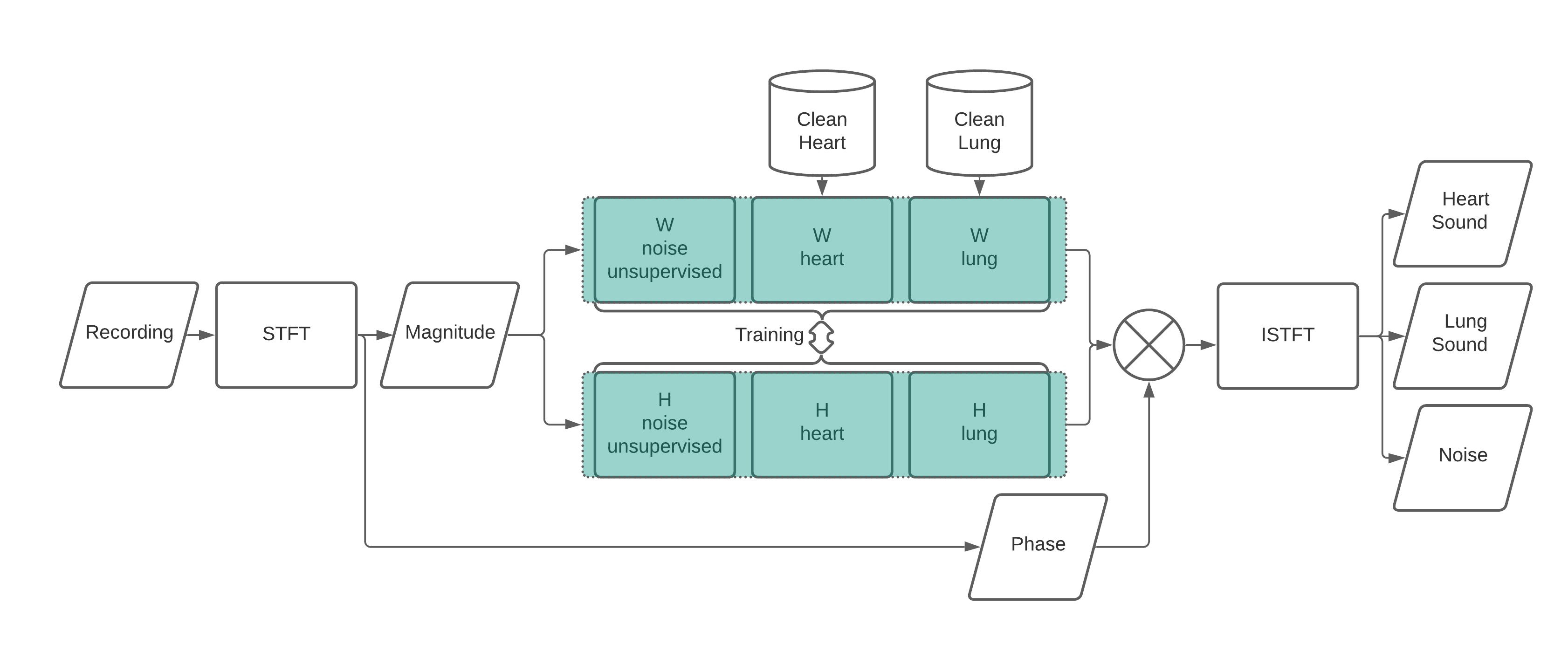} \\
\caption{NMCF Flowchart}
\label{fig:flowchart}
\end{figure*}

However, as obtaining pure heart or lung sounds is not feasible, we propose a modified version of NMCF (Eq.\ref{eq:mnmcf}, Algorithm \ref{alg:mnmcf}, Figure \ref{fig:flowchart}). In this version, datasets of high quality heart, and lung sounds are used in the cost function to enable generalisation of $W_h$ and $W_l$ respectively for sound separation. 

The weighting factors $\lambda_h$ and $\lambda_l$ are determined based on prediction probability of being high quality using automated methods generated in previous works and discussed in \ref{sec:sqi} \cite{grooby2020neonatal}. Additionally, an unsupervised component $W_{un}$ is added to deal with the large variety of noises that are not covered, therefore avoiding these components being assigned to the heart or lung components.

\begin{equation}
\label{eq:mnmcf}
\begin{split}
W_{opt},H_{mopt} & =\min\limits_{W,H}(D(V_m|\hat{W}H_m)\\
&+\sum_{ih=1}^{eh} \lambda^{(ih)}_h D(V_h^{(ih)}|\hat{W_h}H_h^{(ih)})\\
&+\sum_{il=1}^{el} \lambda^{(il)}_l D(V_l^{(il)}|\hat{W_l}H_l^{(il)}))\\
Where: & \quad H_m = [H_{mh};H_{ml};H_{mun}],\\ 
& \hat{W}=[\hat{W}_{h},\hat{W}_{l},\hat{W}_{un}]\\
\end{split}
\end{equation}

\begin{algorithm} 
	\caption{Proposed NMCF}
	\label{alg:mnmcf}
	\LinesNumbered
	$V_m,Phase=stft(audio_m)$\;
	$V_h^{(ih)}=stft(audio_h^{(ih)})$\;
	$V_l^{(il)}=stft(audio_l^{(il)})$\;
	$init \quad H_{mh},H_{ml},H_{mun},H_h^{ih}, H_l^{il}$\;
	$init \quad \hat{W}_{h},\hat{W}_{l}, \hat{W}_{un}$\;
	\For{i=1:maxiter}
	{
	$H_m \leftarrow H_m \otimes \frac{H_{num}(V_m,W,H_m)}{H_{dem}(V_m,W,H_m)}$
    
    $H_h^{ih} \leftarrow H_h^{ih} \otimes \frac{H_{num}(V_h^{ih},W_h,H_h^{ih})}{H_{dem}(V_h^{ih},W_h,H_h^{ih})}$\;
    
    $H_l^{il} \leftarrow H_l^{il} \otimes \frac{H_{num}(V_l^{il},W_l,H_l^{il})}{H_{dem}(V_l^{il},W_l,H_l^{il})}$\;
	
    $W_{h} \leftarrow W_h \otimes 
    \frac{W_{num}(V_m,W_{h},H_{mh})
    +\sum_{ih=1}^{eh}\lambda_h^{eh} W_{num}(V^{(ih)},W_{h},H_h^{(ih)})}
    {W_{dem}(V_m,W_{h},H_{mh})
    +\sum_{ih=1}^{eh}\lambda_h^{eh} W_{num}(V^{(ih)},W_{h},H_h^{(ih)})}$\;
    
    $W_h = normalisation(W_h)$\;
            
    $W_{l} \leftarrow W_l \otimes 
    \frac{W_{num}(V_m,W_{l},H_{ml})
    +\sum_{il=1}^{el}\lambda_l^{el} W_{num}(V^{(il)},W_{l},H_l^{(il)})}
    {W_{dem}(V_m,W_{h},H_{mh})
    +\sum_{il=1}^{el}\lambda_l^{el} W_{num}(V^{(il)},W_{l},H_l^{(il)})}$\;
    
    $W_l = normalisation(W_l)$\;
    
    $W_{un} \leftarrow W_{un} \otimes 
    \frac{W_{num}(V_m,W_{un},H_{mun})}
    {W_{dem}(V_m,W_{un},H_{mun})}$\;
    
    $W_{un} = normalisation(W_{un})$\;
	}
    $W=[\hat{W}_{h},\hat{W}_{l},\hat{W}_{un}]$\;
	$H_{m}=[H_{mh};H_{ml};H_{mun}]$\;
    $mask_h =\frac{\hat{W}_{h}H_{mh}}{WH}$\;
	$mask_l =\frac{\hat{W}_{l}H_{ml}}{WH}$\;
	$V_h = np.multiply(V_m, mask_h)$\;
	$V_l = np.multiply(V_m, mask_l)$\;
	$audio_{heart}=istft(V_h,Phase)$\;
	$audio_{lung}=istft(V_l,Phase)$\;
\end{algorithm}

\begin{table*}[t]
\caption{Signal Quality}
    \centering
    \begin{tabular}{ |P{3.5cm}||P{3cm}|P{3cm}|P{3cm}|P{3cm}|}
    \hline
    \textbf{Method}
    & \textbf{Heart Rate Median Absolute Error (IQR)}
    & \textbf{Heart Signal Quality Mean Improvement (STD)}
    & \textbf{Breathing Rate Median Absolute Error (IQR)}
    & \textbf{Lung Signal Quality Mean Improvement (STD)}\\
    \hline 
   Proposed NMCF
   & 2.51b/10s (4.63b/10s)
   & 0.60 (0.36)
   & 1.38b/10s (1.89b/10s)
   & 0.22 (0.29)\\
    \hline
    NMF by Shah et al. \cite{shah2014blind}
    & 3.12b/10s (4.99b/10s)
    & 0.08 (0.16)
    & 1.82b/10s (2.07b/10s)
    & -0.05 (0.20)\\
    \hline
    NMF by Canadas-Quesada et al. \cite{canadas2017non}
    & 3.46b/10s (4.32b/10s)
    & 0.09 (0.20)
    & 1.57b/10s (2.42b/10s)
    & 0.03 (0.19)\\
    \hline
    \end{tabular}
    \label{tab1:nmfresults}
    \vspace{+1ex}
   \footnotesize{\\IQR=Interquartile range for MAE, STD= Standard deviation, b/10s=beats/breaths per 10 second recording\\
   Heart and lung signal quality of proposed and existing methods were compared to original noisy chest recording, to calculate overall signal quality improvement.}
\end{table*}
\section{Methods}
\label{sec:methods}

\subsection{Data Acquisition and Preprocessing}
The study was conducted at Monash Newborn, Monash Children’s Hospital, a tertiary-level neonatal unit in Melbourne, Australia, and was approved by the Monash Health Human Research Ethics Committee (HREA/18/MonH/471). 
A total of 298 60s recordings from preterm and term newborns were obtained using a digital stethoscope at 16kHz or 44.1kHz sampling frequency. These recordings were then lowpass filtered to avoid aliasing and down-sampled to 4kHz.
Recordings significantly damaged from artifacts making lung and heart sounds impossible to recover were removed. Then 10-second segments containing both heart and breathing sounds were manually extracted. Further details on data acquisition and preprocessing can be found here \cite{zhou2020acoustic,ramanathan2020assessment,grooby2020neonatal}. Total of 68 10-second segments from 60 patients were used in the performance evaluation of the existing and proposed denoising and sound separation methods.

\subsection{Implementation}
Clean heart and lung sound databases containing 20 high quality heart and lung sounds from 18 subjects were assessed by the method in section \ref{sec:sqi}. These database recordings are separate from recordings and patients from the evaluation set. Short-time Fourier transform (STFT) with window length 2048 samples, 75\% overlap and hanning window was used to obtained frequency representation of databases and noisy chest sound recordings. Heart, lung and noise unsupervised signals ($b_h,b_l,b_n$) were represented by 20, 20 and 10 components, sparsity of 0.001, max iteration of 500 and beta loss of 1 corresponding to Kullback-Leibler divergence were used for NMCF. 

\subsection{Performance Evaluation}
\subsubsection{Heart Rate and Breathing Rate Error}
A key goal of obtaining high quality heart and lung sounds is to achieve accurate heart rate and breathing rate estimates, as they are essential in cardio-respiratory health assessment \cite{king2020tools,jain2019neonatal}. Hence, determining the accuracy of these metrics from recordings allow the evaluation signal quality. 

Reference heart rate and breathing rate were obtained from 6 annotators, with the  mean used for comparison. 
Calculated heart rate is initial determined from peak detection of envelope, which is then used for heart segmentation to obtain overall heart rate estimate. Breathing rate is determined from peak detection of 300-450Hz power envelope. More details of calculated heart and breathing rate can be found here \cite{grooby2020neonatal}. 

Median absolute error and interquartile range of proposed and existing methods are shown in Table \ref{tab1:nmfresults}.

\subsubsection{Signal Quality Assessment}
\label{sec:sqi}
Automated signal quality assessment method to classify heart and lung sounds as high and lung quality using dynamic binary classification was developed in the previous work \cite{grooby2020neonatal}. The prediction probability of high quality is used to provide a score from 0 to 1 with regards to signal quality \cite{grooby2020neonatal}. 

\subsubsection{Statistical Analysis}
Statistical tests were performed to determine if proposed NMCF method is significantly better than existing NMF methods. As heart and breathing rate error, and heart and lung signal qualities are not
normally distributed according to Jarque-Bera test, one-sided Wilcoxon signed-ranked test was performed to test significance.

\section{Results}
\label{sec:results}

\begin{figure}[t]
\centering
\includegraphics[scale=0.43,trim={2.5cm 2cm 0cm 0cm}]{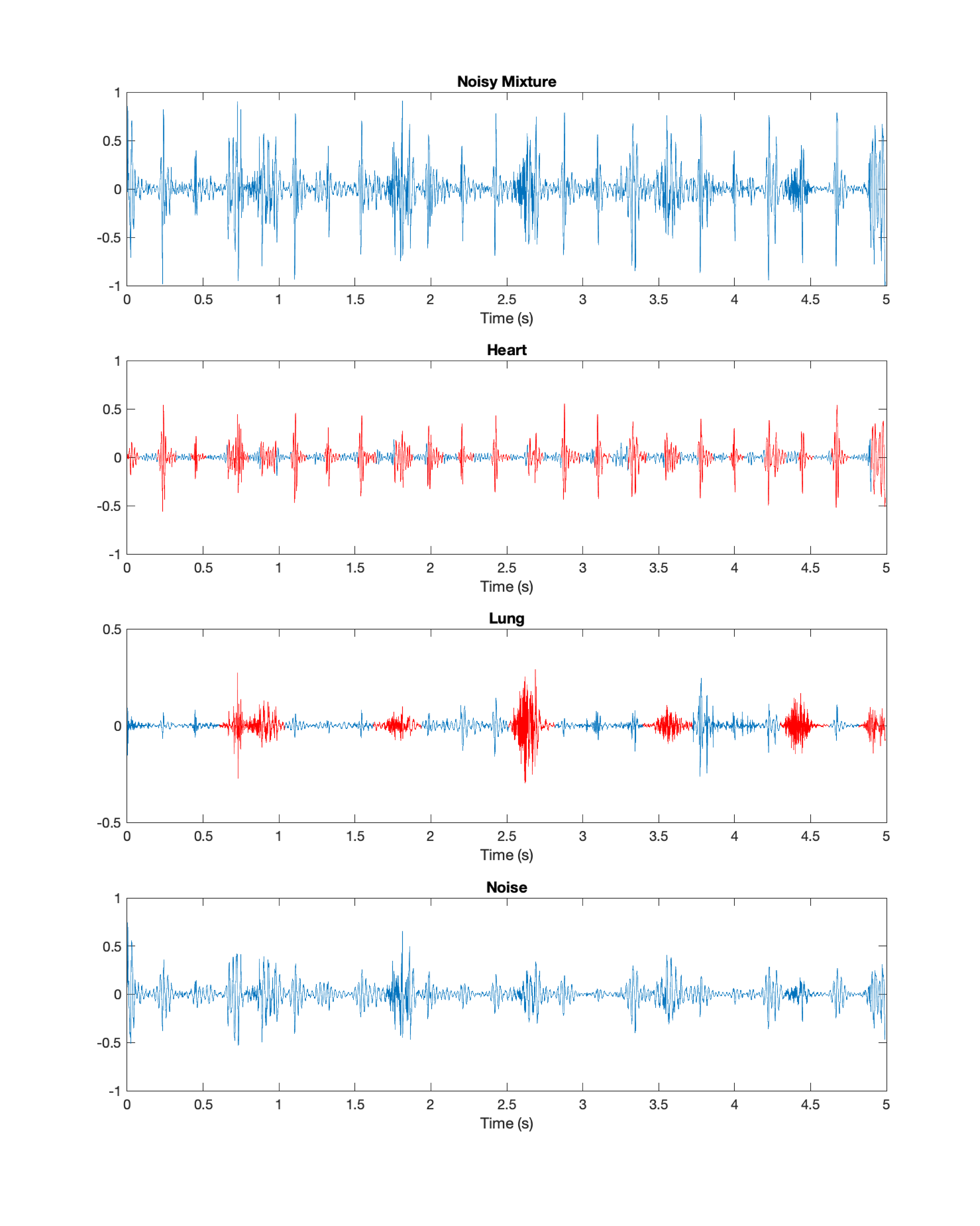} \\
\caption{Example Output from NMCF Method}
\vspace{+1ex}
\footnotesize{In the heart plot, heart beats are highlighted in red using method proposed in \cite{grooby2020neonatal}. In the lung plot, expiration segments are highlighted manually in red, whereas inspiration segments are not highlighted as they are too quiet to be heard or observed easily.}
\label{fig:output}
\end{figure}
In Table \ref{tab1:nmfresults} heart and lung signal qualities of proposed and existing methods were compared to the original noisy recording, to calculate overall signal quality improvement. All methods showed improvement in heart signal quality by 0.60, 0.08 and 0.09 for the proposed and existing methods respectively. However, for lung quality, Shah et al. NMF method saw a decrease by 0.05, whereas it improve by 0.22 and 0.03 in proposed and Canadas-Quesada et al. NMF methods respectively. Overall, proposed method produced significantly better heart and lung signal quality compared to existing methods, with p-values of $2.7\times10^{-15}$, $6.8\times10^{-5}$ respectively. 
The proposed method saw improvement of 0.60b/10s and 0.20b/10s in heart and breathing rate median absolute error compared to existing methods. However, these improvements were not significant, with p-values of 0.10 and 0.14 for heart and breathing rate error improvement respectively. 

Figure \ref{fig:output} shows an example of input noisy mixture recording and the resultant heart, lung and noise components generated from the proposed NMCF method. For heart signal, heart beats can be clearly visualised with minimal noise. While for the lung signal, expiration can be more easily visualised in comparison to the noisy mixture recording, the residual heart beats and noise are still present. The unsupervised noise component appears to successfully aid in the denoising of the heart and lung signals, allowing heart beats and expiration to be more clearly heard and observed as well.

\section{Discussion}
\label{sec:discussion}

As can be seen in Table \ref{tab1:nmfresults}, the proposed NMCF outperforms the two modified existing NMF methods, this was especially prominent in noisy chest sound recordings which high quality heart and lung sound separation was difficult to obtain with the existing methods. The difficulty in dealing with noisy chest sound recordings with the existing methods is due to fact that during the clustering phase, all components are either assigned to heart or lung, meaning no noise components are removed. Hence, as heart sounds are easier to separate, the noise components tend to end up with the separated lung sound. This can explain the decrease in lung signal quality using NMF method proposed by Shah et al., whereas for our proposed method, noisy components that are not closely related to the clean heart and lung sounds are clustered in the noise unsupervised component. 

A potential solution to the existing methods is that during clustering phase, only the top components related heart and lung based on clustering criteria are grouped up, that is the top $b_h$ and $b_l$ components are allocated into heart and lung respectively. The remaining components not clustered are then assumed to be noise and removed. 

However, another issue with the existing methods is the clustering criteria and methodology itself. With the first method, which a reference basis matrix is generated and continually updated during the clustering phase, there is a decent probability that for a recording with a strong noisy component to be incorrectly placed into the reference basis matrix early on. The inclusion of the noisy component in the reference basis matrix then further propagates the error as now further noisy components will be clustered into the reference basis matrix as their correlation criteria has increased. Whereas for the method proposed by Canadas-Quesada et al., the roll-off and temporal correlation criteria are prone to error due to noisy signals. In particular, for temporal correlation, the estimation of S1 and S2 locations within a noisy signal are likely to be inaccurate. 

For low noise recordings, existing methods for heart and lung sound separation perform well, but still outperformed by the proposed method. This is most likely due to the co-factorisation part and number of components. For the existing methods, the recording mixture is blindly decomposed into components and then clustered and into heart and lung components. This blind decomposition means the resultant sub-components may not be directly related to heart and lung, but instead a mixture of heart, lung and low noise content. For the proposed method, the training datasets enable more relevant decomposition to occur. For the NMF by Shah et al., the recording is only decomposed into 20 components which through clustering, results in only 1 heart and 1 lung component for the second stage of NMF. This is too few components to represent complex nature of both heart and lung, which is why in our method 20 components are used to represented heart and lung and another 10 for noise unsupervised component. 

A limitation of the proposed method is the lung quality results, which still contain noise and some remains of heart sound. This is likely due to lung sounds having broad frequency band nature, which overlaps with heart and noise sounds, and having unreliable periodicity, making it harder to denoise and separate. A potential future improvement is the inclusion of a sample noise database into the algorithm, which will assist in the further removal of noise from the lung component. This potential improvement is further supported when analysing the separated noise component, which while containing noise also includes some heart and lung components due to its unsupervised nature. Other potential solutions are the improvement of the clean lung training database for higher quality lung sounds that are less contaminated by heart and noise and parameter optimisation of sparsity, beta loss, window length and percentage overlap of STFT, number of heart, lung and noise unsupervised basis, number of training examples in the databases and relative weighting factor between training examples and the current recording mixture during co-factorisation. 

Another limitation of the proposed method in comparison to the existing methods is the high computational cost of co-factorisation. Two solutions to minimise this effect are introducing an early stopping criteria when the cost function is not decreasing a significant amount and reducing the size of the training examples. Preliminary results suggest that careful selection of 5 heart and lung sounds for the databases may be sufficient. 

Future work on the development of noise and lung spectral correlation criteria may result in a more accurate separation into heart, lung and noise components. Additionally, it may be possible to include these criteria within the framework NMCF to particularly assist in the removal of noise and heart components in the separated lung sound. However, careful consideration into the effects of the cost function and stability of the multiplicative updates is required. 

With regards to the evaluation methods, heart rate and breathing rate error were calculated based on annotator estimations. This method is not the gold standard and in future studies synchronous electrocardiogram will be obtained and used for gold standard heart and breathing rate error calculations. 

Future work is underway to obtain a larger set of high quality heart and lung sounds from newborns, which can be used to generate artificial mixtures which then can be used to more extensively assess denoising and sound separation techniques and optimise the parameters of the proposed NMCF method. 

\section{Conclusion}
\label{sec:conclusion}
This paper focuses on a new approach, based on NMCF in order to obtain high quality heart and lung sounds. The model was trained with high quality heart and lung sounds in parallel with separating sounds from noisy recordings into heart, lung and noise. The training set enabled detailed frequency and temporal aspects of heart and lung to be separated, whereas the parallel training enabled adaptation of the model more specifically to a particular recording and the inclusion of the unsupervised noise component enabled successful denoising of the heart and lung sounds. Overall, this method enables high quality lung and hearts sounds to be generated for analysis. 

\bibliographystyle{IEEEtran}
\bibliography{IEEEabrv,references}

\end{document}